\documentclass[sigconf]{acmart}

\AtBeginDocument{%
  \providecommand\BibTeX{{%
    \normalfont B\kern-0.5em{\scshape i\kern-0.25em b}\kern-0.8em\TeX}}}

\copyrightyear{2021}
\acmYear{2021}
\setcopyright{acmcopyright}
\acmConference[ICTIR '21]{Proceedings of the 2021 ACM SIGIR International Conference on the Theory of Information Retrieval}{July 11, 2021}{Virtual Event, Canada}
\acmBooktitle{Proceedings of the 2021 ACM SIGIR International Conference on the Theory of Information Retrieval (ICTIR '21), July 11, 2021, Virtual Event, Canada}
\acmPrice{15.00}
\acmDOI{10.1145/3471158.3472227}
\acmISBN{978-1-4503-8611-1/21/07}

\usepackage{bm}
\usepackage{ulem}
\usepackage{caption}
\usepackage{graphicx}
\usepackage{subfigure}
\usepackage{amsmath}
\usepackage{booktabs}
\usepackage{threeparttable}
\usepackage{multirow}
\usepackage{enumitem}
\usepackage{xcolor}

\begin{document}
\fancyhead{}

\title{A Discriminative Semantic Ranker for Question Retrieval}

\author{
Yinqiong Cai$^{1, 2}$, Yixing Fan$^{1, 2}$, Jiafeng Guo$^{1, 2}$,
 Ruqing Zhang$^{1, 2}$, Yanyan Lan$^{3}$ and Xueqi Cheng$^{1, 2}$}
\affiliation{
  \institution{
  $^{1}$University of Chinese Academy of Sciences, Beijing, China   \\
  $^{2}$CAS Key Lab of Network Data Science and Technology, Institute of Computing Technology, \\ Chinese Academy of Sciences, Beijing, China\\
  $^{3}$Institute for AI Industry Research, Tsinghua University, Beijing, China\\} 
}
\email{{caiyinqiong18s,fanyixing,guojiafeng,zhangruqing,cxq}@ict.ac.cn, lanyanyan@tsinghua.edu.cn}

\renewcommand{\authors}{Yinqiong Cai, Yixing Fan, Jiafeng Guo, Ruqing Zhang, Yanyan Lan and Xueqi Cheng}
\renewcommand{\shortauthors}{Yinqiong Cai, Yixing Fan, Jiafeng Guo, Ruqing Zhang, Yanyan Lan and Xueqi Cheng}

\begin{abstract}
Similar question retrieval is a core task in community-based question answering (CQA) services. To balance the effectiveness and efficiency, the question retrieval system is typically implemented as multi-stage rankers: The first-stage ranker aims to recall potentially relevant questions from a large repository, and the latter stages attempt to re-rank the retrieved results. Most existing works on question retrieval mainly focused on the re-ranking stages, leaving the first-stage ranker to some traditional term-based methods. However, term-based methods often suffer from the vocabulary mismatch problem, especially on short texts, which may block the re-rankers from relevant questions at the very beginning. An alternative is to employ embedding-based methods for the first-stage ranker, which compress texts into dense vectors to enhance the semantic matching. However, these methods often lose the discriminative power as term-based methods, thus introduce noise during retrieval and hurt the recall performance. In this work, we aim to tackle the dilemma of the first-stage ranker, and propose a discriminative semantic ranker, namely DenseTrans, for high-recall retrieval. Specifically, DenseTrans is a densely connected Transformer, which learns semantic embeddings for texts based on Transformer layers. Meanwhile, DenseTrans promotes low-level features through dense connections to keep the discriminative power of the learned representations. DenseTrans is inspired by DenseNet in computer vision (CV), but poses a new way to use the dense connectivity which is totally different from its original design purpose. Experimental results over two question retrieval benchmark datasets show that our model can obtain significant gain on recall against strong term-based methods as well as state-of-the-art embedding-based methods.
\end{abstract}



\maketitle

\section{Introduction}
Community question answering (CQA) services, such as WikiAnswers, Quora, and Stack Overflow, have grown in popularity in recent years as a platform for people to share knowledge and information. One of the core tasks in CQA is to retrieve similar questions from the archived repository to address user's information needs. In practice, the question retrieval system generally employs multi-stage rankers to balance model complexity and search latency. The first-stage ranker aims to recall a small number of potentially relevant questions from a large repository efficiently. Then, several latter rankers are employed to rerank the initial candidates. Such a multi-stage ranking pipeline has attracted great interest from academia \cite{chen2017efficient, matveeva2006high} as well as industry \cite{liu2017cascade, pedersen2010query}.

However, most existing works on question retrieval mainly focused on building machine learning models for the re-ranking stages, leaving the first-stage ranker to some traditional term-based methods, such as TF-IDF \cite{salton1988term} and BM25 \cite{robertson2009probabilistic}. Specifically, the term-based methods treat each term as a discrete symbolic feature, and represent the questions by bag-of-words (BoW) representations. An inverted index is then built for the corpus on each term and the search process is typically based on the exact matching of 
question terms. However, a major drawback of the term-based methods is the well-known vocabulary mismatch problem~\cite{furnas1987vocabulary}, making the first-stage ranker as a ``blocker'' which prevents the re-rankers from relevant questions at the very beginning. Obviously, this problem would be significantly enlarged on question retrieval due to the sparsity nature of questions.

An alternative of term-based methods is to employ embedding-based methods to enhance semantic matching for the first-stage retrieval. The early effort in this direction dates back to the Latent Semantic Analysis (LSA)~\cite{deerwester1990indexing}. In recent years, with the resurgence of deep learning technique, neural embedding models, from shallow embedding (e.g, word2vec) \cite{vulic2015monolingual, ganguly2015word} to deep contextual embedding (e.g., BERT based models) \cite{karpukhin2020dense, khattab2020colbert}, have been employed for the first-stage retrieval.
Without loss of generality, the embedding-based methods often leverage a dual encoder architecture to compress both users’ questions and historical questions into standalone low-dimensional dense vectors respectively. For example, ColBERT~\cite{khattab2020colbert} is a recently introduced state-of-the-art retrieval model, which employs BERT-based dual encoder architecture to learn contextualized embedding representations for input texts. An approximate-nearest-neighbor (ANN) search is then conducted to retrieve top-k similar questions. With the powerful deep neural networks, these methods are able to learn complex syntactic and semantics of input questions for better question retrieval. 

\begin{figure}[!t]
\centering
\includegraphics[scale=0.35]{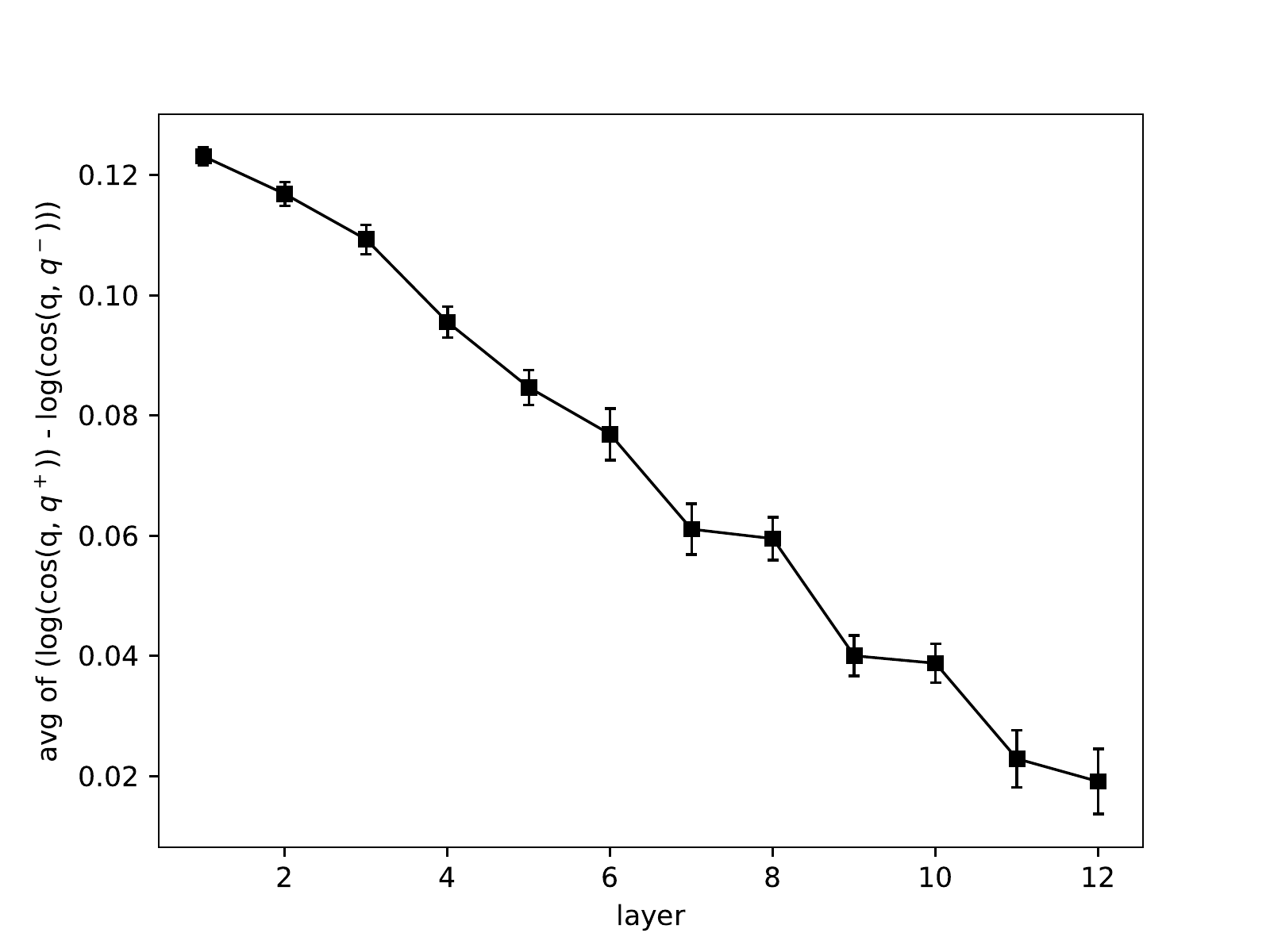}
\caption{
The average of $|log(cos(q, q^+))-log(cos(q, q^-))|$ over all questions in Quora, where $q$ denotes a user question, and $q^+$ and $q^-$ denote the relevant and irrelevant questions of $q$ respectively. The downward trend shows that the discriminative power of representations in BERT decreases when the layer goes deeper.} 
\label{discrimination}                                  
\end{figure}

Despite the significant progress of embedding-based methods for the first-stage retrieval, most existing works focused on learning abstract representations for semantic matching. However, such semantic compression is a double-edged sword, as it may also introduce noise and decrease the discriminative power of representations by dropping detailed symbolic features. For example, we fine-tune the BERT-based dual encoder model on Quora dataset\footnote{https://data.quora.com/First-Quora-Dataset-ReleaseQuestion-Pairs}, and then compute the average difference between the similarity of relevant question pairs and that of irrelevant question pairs by using question representations (i.e., [CLS]) from different BERT layers. We plot this difference against the BERT layer in Figure \ref{discrimination}. As we can see, the difference decreases rapidly as the layer of question representation goes deeper. In other words, high-level abstract representations tend to lose the discriminative power. 

This raises a question to us: is there a way to design an embedding-based ranker that can still keep the discriminative power for high-recall retrieval of questions? To tackle this question, we propose a novel neural embedding model, namely Densely Connected Transformer (DenseTrans), as the first-stage ranker for question retrieval. The DenseTrans model utilizes the Transformer~\cite{vaswani2017attention} architecture to abstract semantic representations for user’s question and historical questions respectively. Specifically, we add dense connections between Transformer layers bottom-up, which help to promote low-level detailed features into high-level representations. In this way, we aim to keep the discriminative power of the dense representations during semantic compression. Note that DenseTrans is inspired by DenseNet \cite{huang2017densely} in computer vision (CV), but poses a new way to use the dense connectivity which is totally different from its original design purpose. In DenseNet, the densely connected layers are introduced to ensure the information flow (e.g., the gradients) between layers in training very deep networks in CV.  While in DenseTrans, we utilize the dense connectivity to retain low-level features in the learned representations to enhance the discriminative power of the semantic ranker. 
Given the learned DenseTrans model, the historical questions in a corpus repository can be pre-encoded and indexed using ANN algorithms \cite{muja2014scalable, zhang2019grip} offline. For online question retrieval, the user’s question is encoded by representation function, and the cosine similarities are computed between user’s question vector and historical question vectors for the first-stage retrieval.

We conduct experiments on two question retrieval benchmark datasets, i.e., Quora and WikiAnswers, to evaluate the effectiveness of our proposed model. Empirical results demonstrate that our DenseTrans model can obtain significant gain on recall against state-of-the-art term-based, embedding-based, and hybrid methods. Meanwhile, DenseTrans also improves the ranking performance in terms of NDCG, MAP and MRR.
We further conduct extensive studies to compare alternative implementations. The results show the importance of dense connectivity on strengthening the low-level features during semantic abstraction to keep the discriminative power of the learned representations.

The remainder of this paper is organized as follows. In Section 2, we introduce the related work to this study. We then describe our proposed method for question retrieval in detail in Section 3. Experimental methodologies and results are presented in Section 4.  In Section 5 we conclude this work and discuss future directions.

\vspace{-0.2 cm}
\section{Related Work}
In this section, we briefly review the most related topics to our work, including question retrieval and first-stage ranking methods.

\subsection{Question Retrieval}
The question retrieval task aims to find similar questions from the archived repository for a new question issued by a user. As a retrieval task, the new user question is taken as a query and the archived historical questions are ranked based on their semantic similarity to the new question.

Similar to other retrieval tasks, the question retrieval task usually employs a multi-stage retrieval pipeline, which requires the search system to firstly retrieve a subset of candidates from the whole collection, and then re-rank the candidates to generate the final results. In practice, the classical retrieval methods, e.g., BM25~\cite{robertson2009probabilistic}, are often applied for the first-stage retrieval, and the re-ranking stages going through quick technique shifts~\cite{robertson1976relevance, liu2011learning, tay2018co}.

Early studies on question retrieval mainly focused on designing effective features to measure the similarities between two questions, such as lexical features, and syntactic features. For example, Wang et al.~\cite{wang2009syntactic} tackled the similar question matching problem using syntactic parsing, while Zhou et al.~\cite{zhou2011phrase} proposed a phrase-based translation model for this task. Although these methods have shown impressive results, they are restricted in their capacity of modeling word sequence information. 

In recent years, along with the development of deep learning technique in information retrieval (IR) community, we have witnessed an explosive growth of research interests on designing neural ranking models for question retrieval tasks. For example, Qiu et al.~\cite{qiu2015convolutional} employed convolutional neural network to encode questions in semantic space. Pang et al.~\cite{pang2016text} evaluated the question similarity from hierarchical levels.
Yang et al.~\cite{yang2019simple} built RE2 model with stacked alignment layers to keep the model fast while still yielding strong performance. Furthermore, many works~\cite{ahasanuzzaman2016mining, gupta2019faq, 10.1145/3397271.3401143} considered the use of different kinds of complementary information, such as question category, Wikipedia concepts and corresponding answers, for the question retrieval task.

\subsection{First-stage Ranking Methods}
In this subsection, we review existing ranking methods for the first-stage retrieval, including term-based, embedding-based, and hybrid methods.

In practice, retrieval systems typically use the term-based methods as the first-stage ranker, such as the vector space model \cite{salton1975vector}, probabilistic model~\cite{robertson2009probabilistic} and language model \cite{ponte1998language}. In the vector space model, queries and documents are taken as bags-of-words while constructing their representation vectors. Next, various scoring functions can be used to calculate the relevance score for each query-document pair. 
These term-based methods form the foundation of modern retrieval systems. However, since they evaluate relevance in the original term space, they easily suffer from the vocabulary mismatch problem. 

In order to tackle the deficiencies of term-based rankers, numerous embedding-based methods~\cite{deerwester1990indexing, gillick2018end, zamani2018neural} have been proposed.
One of the early methods is Latent Semantic Analysis (LSA)~\cite{deerwester1990indexing}. However, LSA is a linear method, which restricts its performance on capturing complex semantic relationships.
With the revival of deep neural networks, the rise of word embeddings stimulates a large amount of works~\cite{mitra2016dual, gillick2018end} exploiting word embeddings to address the vocabulary mismatch problem in retrieval tasks. However, these methods often build the retrieval model based on bag-of-embedded-words ignoring word order features, which are of great importance for text understanding. Recently, neural models have been applied on retrieval tasks maturely, and more sophisticated embedding models, e.g., QA\_LSTM~\cite{tan2015lstm}, ColBERT~\cite{khattab2020colbert}, are proposed. However, due to the loss of detailed low-level features during representation learning process, these models usually have unfulfilling performance over the term-based methods for the first-stage retrieval.

Expecting to enjoy the merits of both, several models~\cite{wei2006lda, mitra2016dual, ganguly2015word} propose to explicitly combine the term-based and embedding-based methods. For example, 
DESM$_{MIXTURE}$~\cite{mitra2016dual} and GLM~\cite{ganguly2015word} linearly combine the scores computed by term-based and embedding-based methods. Santos et al.~\cite{dos-santos-etal-2015-learning} proposed to combine a bag-of-words representation with a distributed vector representation created by a convolutional neural network for retrieving similar questions. These hybrid models slightly improve the performance over term-based methods, but usually along with higher index space occupation and retrieval complexity.

\begin{figure*}[ht]
\centering
\includegraphics[scale=0.5]{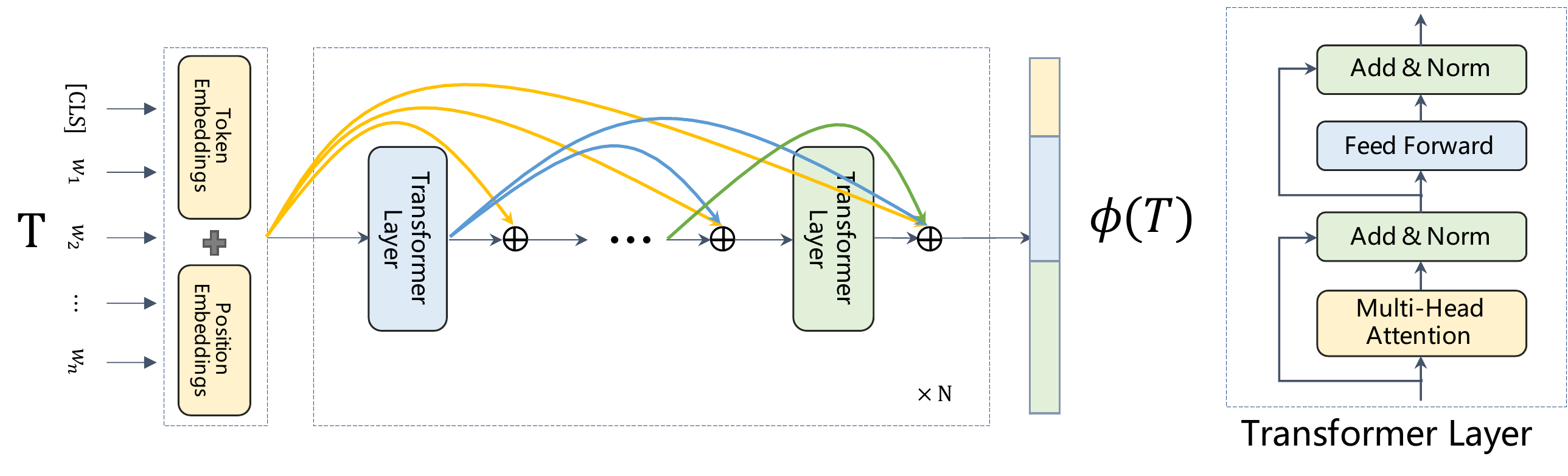}
\caption{The architecture of the Densely Connected Transformer (DenseTrans) model for question retrieval.}
\label{model_DCTM}
\end{figure*}

\section{Our Approach}
In this section, we introduce the DenseTrans model as the first-stage ranker for question retrieval. Section 3.1 discusses the design desiderata. Section 3.2 describes the details of the DenseTrans model. Finally, Section 3.3 presents the model training method.

\subsection{Design Desiderata}
The first-stage ranker for question retrieval aims to recall a list of potentially relevant historical questions $q^h$ from a large repository with respect to a new user's question $q^u$. To satisfy the efficiency requirement of the first-stage ranker, the most popular way is to pre-compute the representations of all the questions in a repository and index them properly offline. During online retrieval, the user's question representation is compared against the historical question representations using some simple relevance functions to quickly recall potentially relevant questions. Without loss of generality, such a process could be formulated by the following dual encoder architecture:
\begin{equation}
\text s(q^u, q^h) = \psi (\phi_{1}(q^u), \phi_{2}(q^h)),
\end{equation}
where $\phi_{1}$ and $\phi_{2}$ denote the representation functions for user's question and historical questions respectively, and $\psi$ denotes the relevance scoring function. In practice, $\psi$ is usually implemented by some cheap similarity functions such as dot product or cosine similarity. To enable efficient nearest neighbors search, we use cosine similarity as the implement of $\psi$ as in prior works~\cite{gillick2018end, khattab2020colbert}. So the remaining question is how to define the representation functions. In order to achieve high-recall retrieval, there are two major requirements on the representations that guide us to design the new representation functions in this work.

\begin{itemize}[leftmargin=*]
\item \textbf{Semantic Requirement:} The representations should have strong semantic encoding power so that semantically relevant questions could be recalled at this stage. In classical term-based methods \cite{salton1975vector, robertson2009probabilistic}, $\phi$ refers to the BoW representation which encodes both user's question and historical questions into sparse symbolic vectors. As a result, only syntactic matching is conducted between user's question and historical questions and recall is significantly hurt due to the severe vocabulary mismatch problem over short texts. To solve this problem, embedding-based methods leverage an embedding function $\phi$ to compress both user's question and historical questions into low-dimensional dense vectors, which can enhance semantic matching by mapping different words into ``latent terms''. In recent years, it has shown that \textit{contextual representation} (e.g., ELMo~\cite{peters-etal-2018-deep}, BERT~\cite{devlin2018bert}) can achieve significantly better performance in many NLP tasks than those non-contextual ones (e.g., word2vec) due to their stronger semantic encoding power. Therefore, we propose to leverage the Transformer architecture~\cite{vaswani2017attention}, the most popular contextual representation learning model, to learn better semantic representations for question retrieval.

\item \textbf{Discriminative Requirement:} The representations should be able to keep discriminative features so that irrelevant questions could be effectively filtered at this stage. Although embedding-based methods could enhance semantic matching through compression, it also introduces noise and decreases the discriminative power of representations by dropping detailed symbolic features. This problem becomes more severe in deep neural embedding models due to the deep abstraction. One solution to this problem is to promote those discriminative low-level features into the abstract high-level representations. This requires us to add short-cuts between layers in conventional neural embedding models. In fact, there have been two types of architectures, i.e., ResNet~\cite{he2016deep} and DenseNet~\cite{huang2017densely}, that successfully add short paths between layers. The ResNet combines the features through \textit{summation} before they are passed into higher layers. As a result, it is still difficult to keep the discriminative power since the low-level features are blended with high-level features. In contrary, the DenseNet leverages \textit{concatenation} to pass the low-level features layer-by-layer. In this way, the low-level features could be kept unchanged and be successfully promoted into the final representations. Therefore, in this work, we propose to take the dense connections to enhance the discriminative power of the semantic representations.
\end{itemize}

\subsection{Densely Connected Transformer Model}
Based on the above design desiderata, we introduce the densely connected Transformer model (DenseTrans) as the first-stage ranker for question retrieval. As is shown in the Figure~\ref{model_DCTM}, DenseTrans consists of three major components, namely the input representation layer, the Transformer encoding layer, and the dense connectivity. In the following, we will describe each component in detail.

\subsubsection{{\bfseries The Input Representation Layer.}} 
The input of DenseTrans model is a sequence of tokens $T = \{w_1, w_2, \cdots, w_n\}$, where $n$ is the length of $T$. We add a special token `[CLS]' before $w_1$ as the pooling token. 
To capture the word order features, we follow existing works \cite{devlin2018bert} to inject absolute positional information to a representation vector, and combine it with the token embedding to form the vector of each token. Here, we learn the position embeddings with the same dimension as token embeddings from scratch as in \cite{vaswani2017attention}. Then, the two embeddings are added up as output of the input representation layer.

\subsubsection{{\bfseries The Transformer Encoding Layer.}} \label{Transformer Encoding Layer}
Here, we take the widely successful Transformer architecture~\cite{vaswani2017attention} as the implementation of the encoding layer. As shown in Figure \ref{model_DCTM}, $N$ Transformer encoding layers are stacked to compute the contextual representation of text $T$. It builds on the self-attention layer, which attends to all positions of the previous layer. In this way, it captures global contextual information more directly.

There are two sub-layers in each encoding layer. The first sub-layer is a multi-head attention structure. The multi-head attention projects the input sequence to query, key, and value inputs of the scaled dot-product attention for each attention head. Then, the results of each attention head are concatenated and projected to the output. Specifically, given a matrix of $n$ query vectors $\mathbf{Q} \in \mathbb{R}^{n \times d}$, keys $\mathbf{K} \in \mathbb{R}^{n \times d}$ and values $\mathbf{V} \in \mathbb{R}^{n \times d}$, the calculation is conducted as follows:
\begin{equation}
\begin{aligned} 
\operatorname{MultiHead}(\mathbf{Q}, \mathbf{K}, \mathbf{V}) &= \operatorname{Concat}(\text{head}_{1}, \cdots, \text{head}_{\mathbf{h}}) \mathbf{W}^{O}, \\
\text{where head}_{\mathbf{i}}  &= \operatorname{Attention}(\mathbf{Q} \mathbf{W}_{i}^{Q}, \mathbf{K} \mathbf{W}_{i}^{K}, \mathbf{V} \mathbf{W}_{i}^{V}), \\
\text{Attention}&(\mathbf{Q}, \mathbf{K}, \mathbf{V}) = \operatorname{softmax}(\frac{\mathbf{Q} \mathbf{K}^{T}}{\sqrt{d}}) \mathbf{V},
\end{aligned}
\end{equation}
where $d$ is the dimension size, $\mathbf{W}_{i}^{Q} \in \mathbb{R}^{d \times d/H}$, $\mathbf{W}_{i}^{K} \in \mathbb{R}^{d \times d/H}$, $\mathbf{W}_{i}^{V} \in \mathbb{R}^{d \times d/H}$ and $\mathbf{W}^{O} \in \mathbb{R}^{d \times d}$ are the parameter matrices to be learnt.
In this situation, we use its self-attention variant, so $\mathbf{Q} = \mathbf{K} = \mathbf{V}$.
The second sub-layer is a position-wise fully connected feed-forward network, which consists of two linear transformations with a ReLU activation in between \cite{vaswani2017attention}, 
\begin{equation}
\operatorname{FFN}(\mathbf{x}) = \max (0, \mathbf{x} \mathbf{W}_{1} + \mathbf{b}_{1}) \mathbf{W}_{2} + \mathbf{b}_{2}.
\end{equation}
Besides, there is a residual connection \cite{he2016deep} around each of the two sub-layers, and a layer normalization \cite{ba2016layer} is followed.

We use $SA(\cdot)$ to denote the process of each encoding layer:
\begin{equation}
\mathbf{E}^{\ell} = SA_{\ell}(\mathbf{E}^{\ell-1}), \label{con:encoding layer}
\end{equation}
where $\mathbf{E}^{\ell}$ denotes the output of the $\ell$-th encoding layer, and $\mathbf{E}^{0}$ is the output of representation layer. We take the output at `[CLS]' of the last encoding layer as the text representation vector.

\subsubsection{{\bfseries The Dense Connectivity.}} 
In order to alleviate the information loss, especially the detailed low-level features, we add dense connections between each Transformer layer. The dense connectivity is inspired by the DenseNet model. 
However, compared with DenseNet, we do not use the \textsl{transition layers} and the \textsl{batch normalization} since the DenseTrans only uses a few layers.
As a result, the direct connections from any layer to all subsequent layers can further improve information flow from lower layers to the upper, so that the representation vectors can retain the detailed low-level features and abstract high-level features simultaneously. That is, the $\ell$-th Transformer encoding layer receives the output matrices of all the preceding layers as input, then the equation \eqref{con:encoding layer} can be re-written as follows:
\begin{equation}
\mathbf{E}^{\ell} = SA_{\ell}([\mathbf{E}^0; \mathbf{E}^1; \cdots; \mathbf{E}^{\ell-1}]),
\end{equation}
where $[\mathbf{E}^0; \mathbf{E}^1; \cdots; \mathbf{E}^{\ell-1}]$ is the concatenation of the output matrices produced by representation layer and Transformer encoding layers $1, 2, \cdots, \ell-1$.

\subsection{Model Training}
During training, we utilize cross entropy loss to optimize all parameters in DenseTrans for ranking task. Firstly, we convert the relevance scores obtained by model $s$ through softmax function:
\begin{equation}
s^{*}(q^u, {q^h}^{+}) = \frac{\exp(s(q^u, {q^h}^{+}))}{\sum_{{q^h}^{\prime} \in \bm{Q}} \exp ( s(q^u, {q^h}^{\prime}))} ,
\end{equation}
where \textbf{\textsl{Q}} denotes the question collection in the whole repository. 
In practice, for each similar question pair, denoted by $(q^u, {q^h}^{+})$ where $q^u$ is the user's question and ${q^h}^{+}$ is one of relevant historical questions of $q^u$, we approximate \textbf{\textsl{Q}} by including ${q^h}^{+}$ and other $J$ negative questions. The negative questions come from two source. A part of them are sampled from the retrieval results of a strong heuristic unsupervised model \cite{huang2013learning}, and the sampling strategy depends on the learning datasets, which can be found in section~\ref{sampling_strategy}. Other negative questions are the questions in the same mini-batch.
Then, all parameters are learned to maximize the scores of similar question pairs across the whole training set $\mathcal{D}$. That is, we minimize the following loss function:
\begin{equation}
\mathcal{L}(\mathcal{D}, \Theta) = - \log \prod_{(q^u, {q^h}^{+})} s^{*}(q^u, {q^h}^{+}),
\end{equation}
where $\Theta$ denotes all parameters in DenseTrans. The optimization is done with standard backpropagation.

\section{Experiments}
In this section, we conduct experiments to demonstrate the effectiveness of our proposed model.

\subsection{Datasets Description}\label{sampling_strategy}
Firstly, we introduce the datasets used in our experiments and show the statistical information of them in Table \ref{tab:statistics}.

\textbf{Quora dataset:}\quad The Quora Question Pairs (QQP) dataset is published for paraphrase identification task. We adapt it to question retrieval task. Specifically, we take all questions in this dataset to form a collection, which includes about 538K questions. For each question in the training set with at least one paraphrase question, we use it as a training question and take its paraphrase questions as positive examples. After processing, we get 79.6K training questions, and each of them has 1.69 positive examples averagely. For the development set, we do the same process and get approximately 13K development questions. Since the test set labels are not released, we split the obtained development questions to construct the dev and test set, with each contains 6.5K questions. We filter these questions to ensure they have no overlap with training questions to avoid data leakage. For model training, we retrieve top-100 candidate questions for each training question using BM25 and filter corresponding positive examples, then sample from the results as negative examples for training the ranking model.

\textbf{WikiAnswers dataset:}\quad The WikiAnswers Paraphrase corpus \cite{fader2013paraphrase} contains over 2M questions and the average number of positive examples is 12.85. 
These similar question pairs are gathered from WikiAnswers\footnote{http://wiki.answers.com/}. 
For training efficiency, we sample 100K/5K/5K questions from this dataset without overlapping as the train/dev/test set. Compared with the QQP dataset, each question in WikiAnswers has more relevant questions, thus, its negative sampling strategy is different from QQP dataset. Concretely, we use BM25 to retrieve top-500 results for each question in training set and filter corresponding positive examples, then randomly sample from the results as negative examples for model training. In fact, we carry out a pilot experiment that adopts the same sampling strategy as QQP dataset, but the performance is not ideal. We speculate that it is because there are more positive examples in WikiAnswers and the negative examples need to be more diverse correspondingly. We further try different sampling strategies, and the experimental results and analysis can be found in section~\ref{sampling}.

\begin{table}[!t]
  \caption{Statistical information of datasets.}
  \label{tab:statistics}
  \begin{tabular}{ccccc}
    \toprule
    Dataset & \#q(collection) & \#q(train) & \#q(dev) & \#q(test)\\
    \midrule
    \texttt{Quora} & 537,920 & 79,641 & 6,520 & 6,520\\ 
    \texttt{WikiAnswers}& 2,539,836 & 100,000 & 5,000 & 5,000\\
    \bottomrule
  \end{tabular}
\end{table}

\subsection{Experimental Setup}
\subsubsection{{\bfseries Baseline Methods}}
We consider three types of methods for comparison, including term-based models, embedding-based models, and hybrid models. 

\noindent \textbf{Term-based models:}
\begin{itemize}[leftmargin=*]
\item \textbf{BM25:} BM25 \cite{robertson2009probabilistic} is a probabilistic retrieval model, which usually as a strong baseline for retrieval tasks. The key hyper-parameters are selected by heuristically search.
\item \textbf{RM3:} RM3~\cite{lavrenko2017relevance} is a state-of-the-art pseudo-relevance feedback model to alleviate the vocabulary mismatch problem of term-based methods. We choose the number of feedback documents, feedback terms and feedback coefficient by heuristically search.
\end{itemize}

\noindent \textbf{Embedding-based models:}
\begin{itemize}[leftmargin=*]
\item \textbf{DESM:} DESM~\cite{mitra2016dual} utilizes neural word embeddings as a source of evidence to compute the relevance score by aggregating the cosine similarities across all the question-question word pairs. We choose the DESM$_{\mathit{IN-OUT}}$ version as it has the best performance~\cite{mitra2016dual}, and use the released in and out word embeddings\footnote{http://research.microsoft.com/projects/DESM} for our experiments. 
\item \textbf{DualEncoder}: DualEncoder~\cite{gillick2018end} uses an average over word embeddings to represent the input questions. The experiment uses a multi-task setup, including in-batch cross-entropy, in-batch sampled softmax and in-batch triplet. We implement it based on Pytorch\footnote{https://pytorch.org/} since there is no publicly available codes.
\item \textbf{PV-DBOW:} PV-DBOW~\cite{le2014distributed} learns text embeddings by estimating a language model at the whole text level. We directly use the released model\footnote{https://github.com/jhlau/doc2vec} (trained on wikipedia data) by Lau et al.~\cite{lau2016empirical} to infer question representations in our datasets. For inferring stage, the initial learning rate is tuned between 0.01 and 0.1, and the training epoch for new texts is set to 100.
\item \textbf{QA\_LSTM:} QA\_LSTM~\cite{tan2015lstm} uses BiLSTM and max pooling to construct text representations, and relies on a MLP to calculate the matching scores. We implement it based on Pytorch since there is no publicly available codes.
\item \textbf{SNRM:} SNRM~\cite{zamani2018neural} learns high-dimensional sparse representations for input texts, then uses dot product to calculate the matching scores. We use the model code published by authors \footnote{https://github.com/hamed-zamani/snrm} for our experiments. 
\item \textbf{ColBERT:} ColBERT~\cite{khattab2020colbert} is a recently introduced state-of-the-art model, which is specifically designed for the first-stage retrieval. It employs a cheap interaction function, i.e., a term-based MaxSim, to model the fine-grained matching signals. We use the model code published by authors \footnote{https://github.com/stanford-futuredata/ColBERT} for our experiments. 
\end{itemize}

\noindent \textbf{Hybrid models:}
\begin{itemize}[leftmargin=*]
\item \textbf{BOW-CNN:} The BOW-CNN model~\cite{dos-santos-etal-2015-learning} computes two partial similarity scores: $s_{bow}(q_1, q_2)$ for the BOW representations and $s_{conv}(q_1, q_2)$ for the CNN representations. Finally, it combines the two partial scores to create the final score $s(q_1, q_2)$.
\item \textbf{DESM$_{\mathit{MIXTURE}}$:} The DESM$_{\mathit{MIXTURE}}$ \cite{mitra2016dual} combines DESM with BM25 with a hyper-parameter $\alpha$ to adapt to the retrieval tasks on large-scale collections. We perform a parameter sweep between 0 and 1 at intervals of 0.02.
\end{itemize}

\renewcommand{\arraystretch}{1.1}
\begin{table*}[ht]
\large
  \centering
  \fontsize{8.5}{9}\selectfont
  \begin{threeparttable}
  \caption{Performance of our proposed model and baselines. The highest value for every column is highlighted in bold and all the statistically significant (p < 0.05) improvements over the BM25, BOW-CNN and DESM$_{\mathit{MIXTURE}}$ baseline are marked with the asterisk $\ast$, $\dag$ and $\ddag$ respectively.}
  \label{tab:performance_comparison}
    \begin{tabular}{lcccccccc}
    \toprule
    \toprule
    \multirow{2}*{\textbf{Method}}&
    \multicolumn{4}{c}{\textbf{Quora}}&\multicolumn{4}{c}{\textbf{WikiAnswers}}\cr
    \cmidrule(lr){2-5} \cmidrule(lr){6-9}
    &\textbf{Recall@100}&\textbf{MRR@100}&\textbf{MAP@100}&\textbf{NDCG@10}&\textbf{Recall@100}&\textbf{MRR@100}&\textbf{MAP@100}&\textbf{NDCG@10}\cr
    \midrule
    \midrule
    \textbf{BM25} & 0.9275 & 0.5171 &0.5074&0.5565 &0.8053  &0.6845 &0.4821 &0.4649 \cr
    \textbf{RM3} &0.9291 &0.5173 &0.5078 &0.5575  &$0.8138^{\ast}$   &0.6792  & $0.4914^{\ast}$  &0.4689  \cr
    \midrule
    \textbf{DESM} &0.2096 &0.0435 &0.0417 &0.0471 &0.1043  &0.0633 &0.0142 &0.0211 \cr
    \textbf{DualEncoder} &0.7900 &0.4393 &0.4303 &0.4825 &0.5696 &0.6007 &0.3728 &0.3643 \cr
    \textbf{PV-DBOW} &0.7061 &0.3351 &0.3250 &0.3588 &0.6187  &0.6289 &0.3642 &0.3833 \cr
    \textbf{QA\_LSTM}&0.4686 &0.2286 &0.2216 &0.2452 &0.5261  &0.6121 &0.3021 &0.3369 \cr
    \textbf{SNRM} &0.4223 &0.2470 &0.2388 &0.2607 &0.2498 &0.4560 &0.1482  &0.2097 \cr
    \textbf{ColBERT} &$0.9436^{\ast}$  &0.5184  &0.5086  &0.5608  &0.7863  &0.6852  &0.4807  &0.4586  \cr
    \midrule
    \textbf{BOW-CNN} &0.9322  &$0.5310^{\ast}$  &$0.5226^{\ast}$  &$0.5712^{\ast}$  &0.7923  &$0.7018^{\ast}$  &$0.4945^{\ast}$  &$0.4854^{\ast}$  \cr
    \textbf{DESM$_{\mathit{MIXTURE}}$} &$0.9347^{\ast}$ &$0.5352^{\ast}$ &$0.5262^{\ast}$ &$0.5738^{\ast}$ &$0.8031^{\dag}$  &$0.7010^{\ast}$ &$0.4957^{\ast}$ &$0.4792^{\ast}$ \cr
    \midrule
    \textbf{DenseTrans} &${\textbf{0.9707}}^{{\ast} \dag \ddag}$ &${\textbf{0.5483}}^{\ast \dag \ddag}$ &${\textbf{0.5394}}^{\ast \dag \ddag}$ &${\textbf{0.5942}}^{\ast \dag \ddag}$ &${\textbf{0.8362}}^{\ast \dag \ddag}$ &${\textbf{0.7309}}^{\ast \dag \ddag}$ &${\textbf{0.5238}}^{\ast \dag \ddag}$ &${\textbf{0.4933}}^{\ast \dag \ddag}$ \cr
    \bottomrule
    \bottomrule
    \end{tabular}
    \end{threeparttable}
\end{table*}

\subsubsection{{\bfseries Evaluation Metrics}}
As a first-stage ranker, we mainly focus on the capability to recall as many potentially relevant historical questions as possible, so we use recall of top 100 ranked questions (Recall@100) as the main criterion. The recall metrics of other depths (e.g., Recall@10, Recall@20 and Recall@50) are also reported in section~\ref{Different Threshold}.
In addition, we report three other standard evaluation metrics for ranking tasks as previous works, i.e., mean reciprocal rank of top ranked 100 questions (MRR@100), mean average precision of top 100 retrieved questions (MAP@100), and normalized discounted cumulative gain of top 10 ranked questions (NDCG@10).

\subsubsection{{\bfseries Parameter Settings and Implementation Details}}
We use the Anserini\footnote{http://anserini.io/} toolkit \cite{yang2017anserini}, a popular open-source Lucene search engine, to obtain BM25 retrieval results for negative examples sampling. The key hyper-parameters of BM25 are tuned to $k_1 = 3.44$ and $b = 0.87$.
Our models are implemented with PyTorch framework. Note that the DenseTrans model does not utilize any pre-trained transformer weights (e.g., BERT) due to the mismatch of hidden size owing to the dense connections.
Here, the token embeddings are initialized by the 300-dimension pre-trained FastText \cite{bojanowski2017enriching} word vectors and updated during the training process. The out-of-vocabulary (OOV) words are randomly initialized by sampling values uniformly from $(-0.2, 0.2)$. Other parameters are initialized by the default initialization functions in PyTorch. 
All questions in the datasets are truncated or padded to 30 words. The number of Transformer encoding layers and parallel attention heads are set to $N=3$ and $H=6$ respectively. We use a dropout \cite{srivastava2014dropout} rate of $0.1$ on all encoding layers.

We use the Adam optimizer \cite{kingma2014adam} with $\beta_1 = 0.9$, $\beta_2 = 0.98$ and $\epsilon = 10^{-8}$. A scheduler is created with the learning rate decreasing linearly after a linearly increasing process during the warmup period. We set the number of warmup steps to $10\%$ of the total training steps. The batch size is set to 32, and we run all the experiments on Tesla K80 GPUs. For all the models, the hyper-parameters are tuned with the dev set. We pick the model that works best on the dev set, and then evaluate it on the test set. 

We employ an off-the-shelf library for large-scale vector search, namely faiss~\cite{johnson2019billion} from Facebook~\footnote{https://github.com/facebookresearch/faiss}. For our faiss-based implementation, we use an IVFPQ index (“inverted file with product quantization”). For the index constructing, we set the number of partitions to 2000, and divide each embedding into 16 sub-vectors, each encoded using one byte. For online serving, we only search the nearest 10 partitions for the top-100 matches when a question embedding is coming.

\subsection{Main Evaluation Results}
This section presents the performance results of different retrieval models over the two benchmark datasets. A summary of results is displayed in Table \ref{tab:performance_comparison}.

According to the results, BM25 is a strong term-based method for the first-stage retrieval, which achieves good performances on both datasets. The RM3 model only obtains slightly better performance than BM25, and most improvements are not significant. The results indicate that the pseudo-relevance feedback technique might not be that effective on short text retrieval as it has shown on long document retrieval~\cite{DBLP:conf/cikm/GuoFAC16a}. 

For the shallow embedding models such as DESM, DualEncoder and PV-DBOW, we can observe that: 1) A direct use of DESM obtains extremely poor results, which is in consistency with previous work~\cite{mitra2016dual}. The possible reason is that the DESM features are very susceptible to false positive matches under non-telescoping setting~\cite{mitra2016dual}.
2) The DualEncoder with a multi-task setup is more effective than DESM. This is reasonable since DualEncoder is a supervised method which directly optimizes the embeddings towards the question retrieval, while DESM aggregates word embeddings learned in an unsupervised manner. Moreover, the DualEncoder method learns the model with the in-batch loss function, which makes the training more consistent with the inference for the first-stage retrieval.
3) As for PV-DBOW, we can see it outperforms the DESM with a large margin. This maybe that it directly learns a paragraph vector by predicting each word, which can better capture the global semantics of questions. Moreover, the PV-DBOW achieves better performance on WikiAnswers than the DualEncoder. A possible reason maybe that the PV-DBOW model maintains the discriminative power by predicting each word during learning the text representation, which makes it more robust on the noisy dataset (i.e, the WikiAnswers).

For the deep embedding methods, we can find that: 1) The performance of QA\_LSTM is far behind BM25. The results demonstrate that by simply learning high-level abstraction of questions, the QA\_LSTM model is prone to recall non-relevant questions due to the missing of detailed low-level features.
2) The SNRM model, which is designed for the first-stage retrieval, obtains relatively poor results as compared with BM25 either. The possible reason is that SNRM is specially designed for the ad-hoc retrieval, where the documents are usually lengthy with rich content~\cite{zamani2018neural}.
3) The ColBERT model, a recently introduced state-of-the-art method for the first-stage retrieval, achieves the best performance among the deep embedding-based models. Moreover, it obtains better performance than the two strong term-based models (i.e., BM25 and RM3) on Quora dataset, especially in terms of the recall metric.

For the hybrid methods, we can see that both BOW-CNN and DESM$_{\mathit{MIXTURE}}$ can achieve good performance on both dataset. Especially, DESM$_{\mathit{MIXTURE}}$ improves with a large margin over the basic DESM by combining with a term-based method. This demonstrates that the fine-grained term matching signals are very beneficial for question retrieval. Moreover, it can be observed that both BOW-CNN and DESM$_{\mathit{MIXTURE}}$ outperform BM25 significantly in terms of ranking metrics (i.e., MRR@100, MAP@100, and NDCG@10) on both datasets. As for the recall metrics, BOW-CNN and DESM$_{\mathit{MIXTURE}}$ are about on par with BM25. All these results indicate that it is useful to enhance the dense models with low-level matching signals to achieve better retrieval performance.

Finally, our DenseTrans model achieves the best performance on both the Quora and WikiAnswers datasets in terms of all evaluation metrics over all baselines. For example, the improvement of DenseTrans over the best performing baseline method (i.e., DESM$_{\mathit{MIXTURE}}$) is about $3.6\%$ and $3.3\%$ in terms of Recall@100 on Quora and WikiAnswers datasets, respectively. All these results demonstrate the importance to keep both the semantic power as well as the discriminative power in building the first-stage rankers. Besides, it is worth to note that the BOW-CNN and DESM$_{\mathit{MIXTURE}}$ are a linear combination of term-based and embedding-based models, which requires an additional index to support the first-stage retrieval. On the contrary, our DenseTrans model is an end-to-end embedding-based model, which can directly apply a single ANN index for efficient retrieval.

\subsection{Ablation Analysis}
To investigate the impact of architecture design, we compare the performance of  DenseTrans with its variants on the benchmarks.

\subsubsection{Impact of the Dense Connectivity.}
\begin{table}[!t]
\small
\centering
\caption{Results of ablation study on Quora dataset.}
\label{tab:ablation_study}
\setlength{\tabcolsep}{0.7mm}{
\begin{tabular}{ccccc}
\toprule
\textbf{Method} & \textbf{Recall@100} & \textbf{MRR@100} & \textbf{MAP@100} & \textbf{NDCG@10}  \\
\midrule
\textbf{DenseTrans} & 0.9707 & 0.5483 &0.5394 &0.5942\\
\quad  - \textsl{\textbf{TopDense}} & 0.9626 & 0.5275 &0.5188 &0.5743\\
\quad - \textsl{\textbf{AllDense}} & 0.9301 & 0.4714 &0.4629 &0.5211\\
\quad + \textsl{\textbf{Concat}} & 0.9534 & 0.5174 &0.5083 &0.5629\\
\bottomrule
\end{tabular}}
\end{table}

Since we utilize the dense connectivity to enhance the model with detailed low-level features, here, we conduct experiments to verify the effectiveness of the dense connectivity. For this purpose, we introduce several variants of the DenseTrans model. 
Firstly, we remove the dense connections from the last layer, and keep the dense connectivity between low layers. In this way, the top encoding layer takes the outputs of all bottom layers as input to produce the final representation. We denote it the $-\textsl{\textbf{TopDense}}$ as is shown in Table \ref{tab:ablation_study}. From the results we can see that the performance drops slightly compared with original DenseTrans model on the Quora dataset.
Secondly, we remove all the dense connections from the DenseTrans model. In this way, the output representations of the model only keep the highly abstract semantic information. As is shown in Table \ref{tab:ablation_study}, the $- \textsl{\textbf{AllDense}}$ leads to a large performance loss. For example, the Recall@100 has decreased as much as $4.0\%$ compared the DenseTrans model. 
Finally, we further investigate the impact of the detailed low-level features by concatenating the outputs of all layers of the $- \textsl{\textbf{AllDense}}$ to produce the final representation of input text, which we denotes as $+ \textsl{\textbf{Concat}}$ in Table \ref{tab:ablation_study}. The same way to connect layers is also used in~\cite{yin2015multigrancnn, nie2018multi}. From the results, we can see that $+ \textsl{\textbf{Concat}}$ indeed improves the performance on all evaluation metrics over  $- \textsl{\textbf{AllDense}}$. These demonstrate that the detailed low-level features are really important for question retrieval. But, it is worth to note that the performance of $+ \textsl{\textbf{Concat}}$ cannot reach the original DenseTrans model. It indicates the superiority of dense connections, that combine the low-level features before obtaining more abstract representations.

\begin{figure}[!t]
\centering
\includegraphics[scale=0.4]{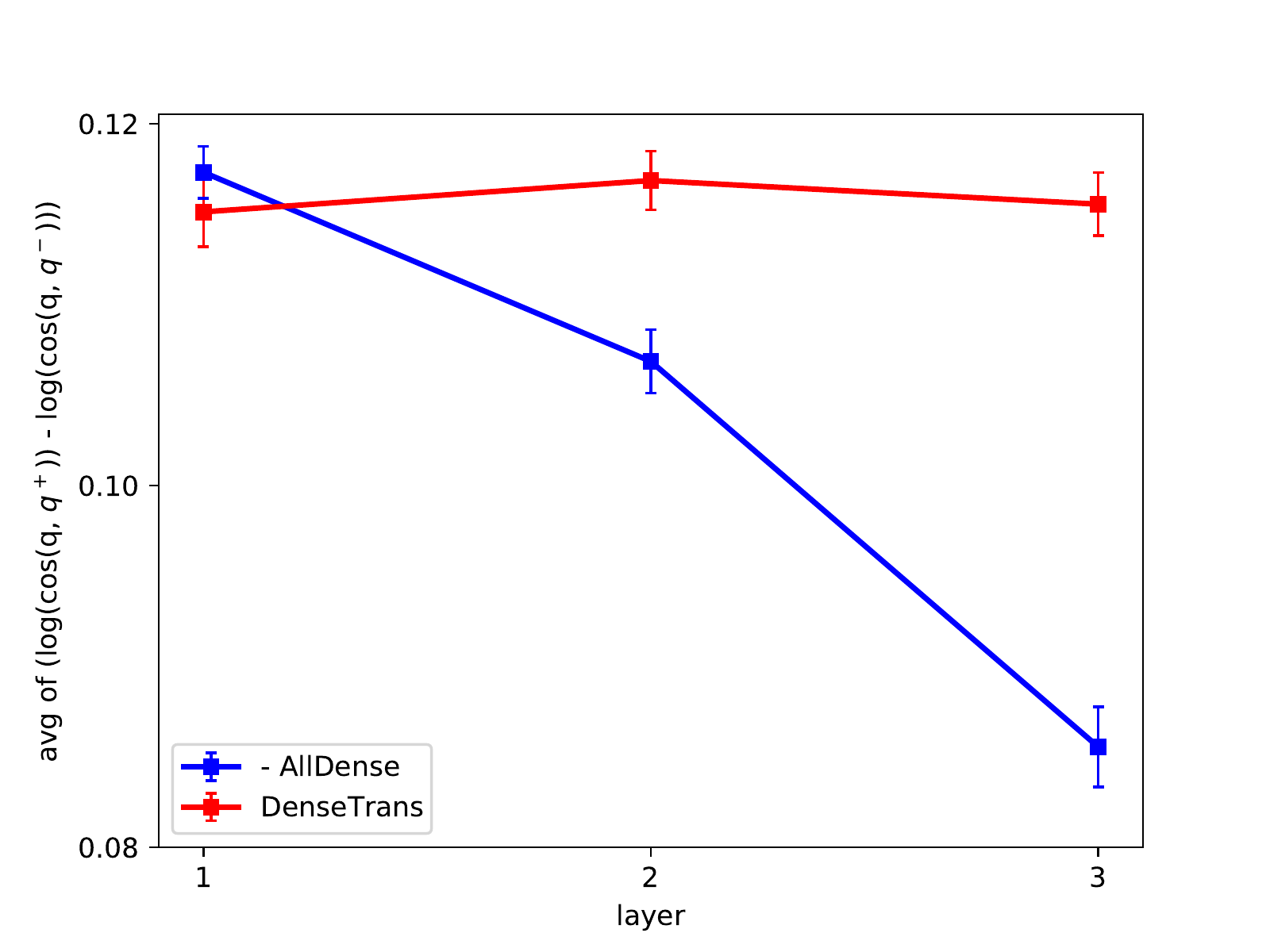}
\caption{The average of $|log(cos(q, q^+))-log(cos(q, q^-))|$ in term of DenseTrans and $- \textsl{\textbf{AllDense}}$  over all questions in Quora dataset.}
\label{fig:nodense_log_mean_distance}
\end{figure}

In order to further check whether dense connections can maintain the discriminative power, we repeat the analysis method in Figure \ref{discrimination} on DenseTrans and $- \textsl{\textbf{AllDense}}$ models and show the results in Figure \ref{fig:nodense_log_mean_distance}. The $- \textsl{\textbf{AllDense}}$ shows the same trend as BERT, that the discriminative power of text representations decreases when the layer goes deeper, while introducing dense connections (DenseTrans) can successfully turn the downward trend. The results prove the effectiveness of dense connections on maintaining the discriminative power.

\begin{figure}[!t]
\centering
\includegraphics[scale=0.4]{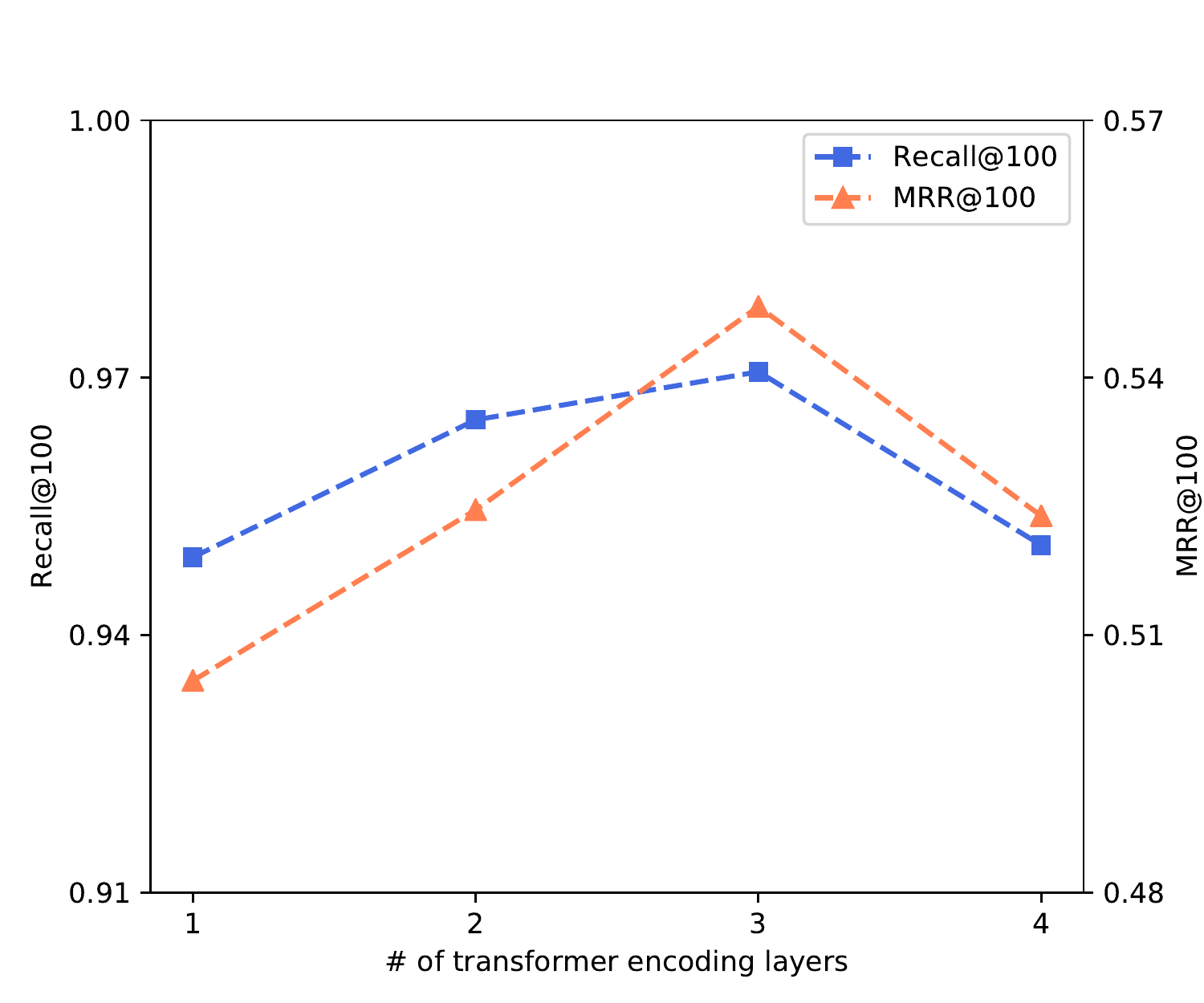}
\caption{The results of Recall and MRR with different number of Transformer layers on Quora dataset.}
\label{fig:layer_numbers}
\end{figure}

\subsubsection{Impact of the Number of Transformer Encoding Layers}
Since our model is built on the stacked Transformer layers. Here, we further study the impact of the number of Transformer layers on representation learning. Specifically, we report the performance results on Quora dataset by stacking 1, 2, 3, and 4 Transformer layers.
The results are shown in Figure \ref{fig:layer_numbers}.
As we can see that the performances of Recall@100 and MRR@100 increase gradually with the number of Transformer layers. Then, the performance decreases sharply if we continue to increase the encoding layer.
A possible reason may be that continuing to increase the encoding layer will dramatically increase the model parameters, and what's more, if the coarse-grained information takes up too much proportion in the text representations, the question pairs matching would be dominated by the high-level abstract features. It would be interesting to further study the balance of the detailed low-level features and the abstract high-level features in the final representations, so we would leave this as a future work.

\begin{figure}[!t]
\centering
\includegraphics[scale=0.4]{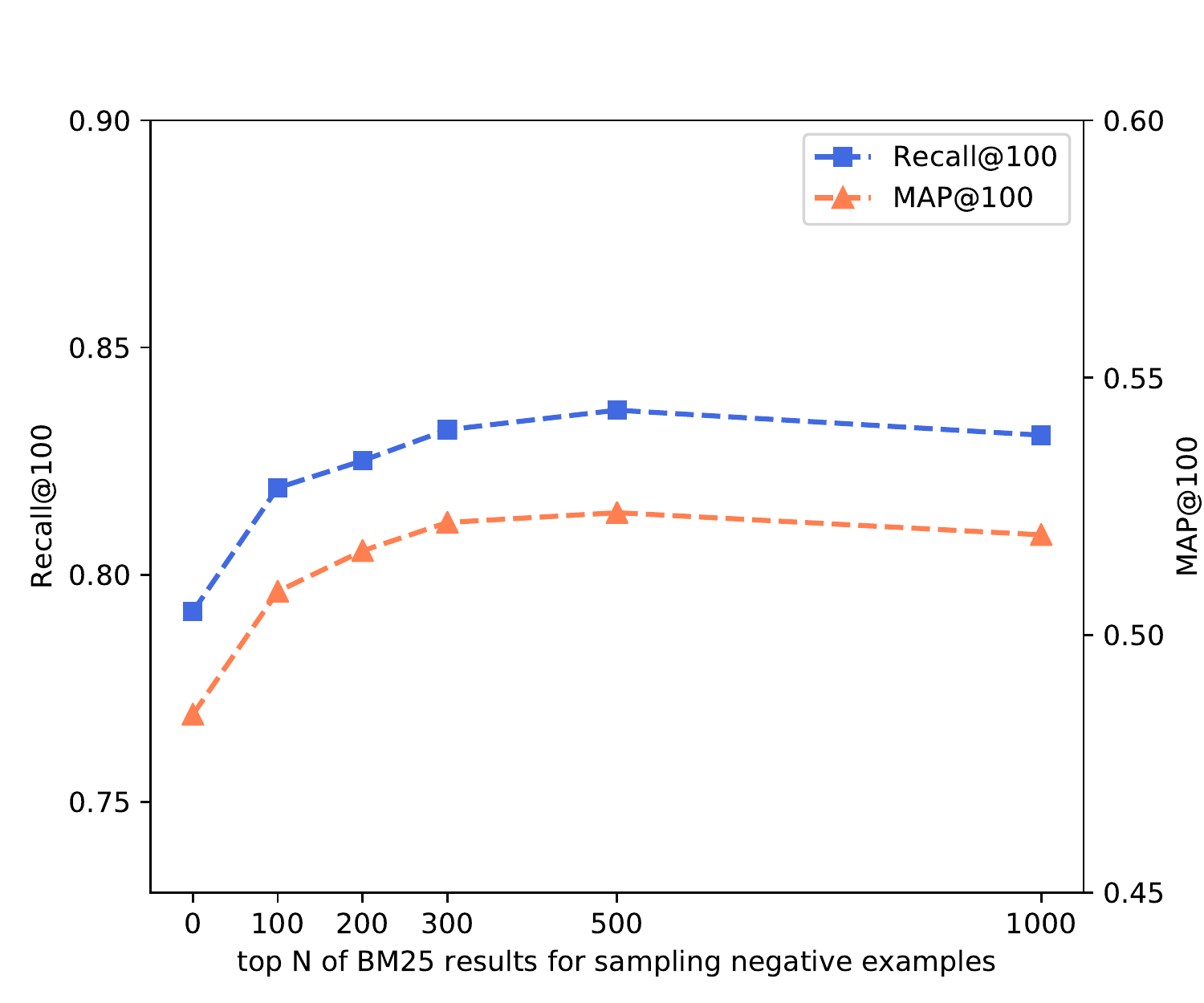}
\caption{The results of Recall and MAP with different sampling strategies for hard negatives on WikiAnswers dataset.}
\label{fig:sampling}
\end{figure}

\subsubsection{Impact of the Sampling Strategies for Hard Negative Examples}\label{sampling}
As described in section \ref{sampling_strategy}, we train the DenseTrans model by sampling hard negative examples from the top ranked results of BM25. In fact, at the beginning, we adopt the same negative sampling strategy as QQP dataset on WikiAnswers dataset, but the performance is not ideal. We speculate that it is because there are more positive examples in WikiAnswers dataset and the negative examples need to be more diverse correspondingly. 
Thus, we conduct a series of experiments on WikiAnswers dataset to investigate the impact of hard negative examples. Specifically, we randomly sample hard negative questions from the top-$N$ ranked results of BM25, where $N$ can be 0, 100, 200, 300, 500, and 1000. It is worth to note that when $N$ is 0, it means that we only use the negatives that come from in-batch and there is no hard negative examples for model training. The results are shown in Figure \ref{fig:sampling}. It can be observed that the DenseTrans model gets better performance on Recall@100 and MAP@100 metrics along with the increase of $N$. Then, the performance decreases slightly if we continue to increase the $N$. The DenseTrans achieves best performance when $N$ equals to 500.
This may be that the larger the $N$ is, the more diverse the sampled negative examples are. At the same time, when $N$ is too large, the negatives are not hard enough. Thus, the $N$ is a balance between the hardness and diversity of the negative examples. Besides, it is worth to note that the performance has a sharply increasing when $N$ is set from 0 to 100. It shows that hard negative examples are essential for the first-stage ranker training, which is consistent with previous works~\cite{luan2020sparse, karpukhin2020dense}.

\begin{figure}[!t]
\centering
\includegraphics[scale=0.43]{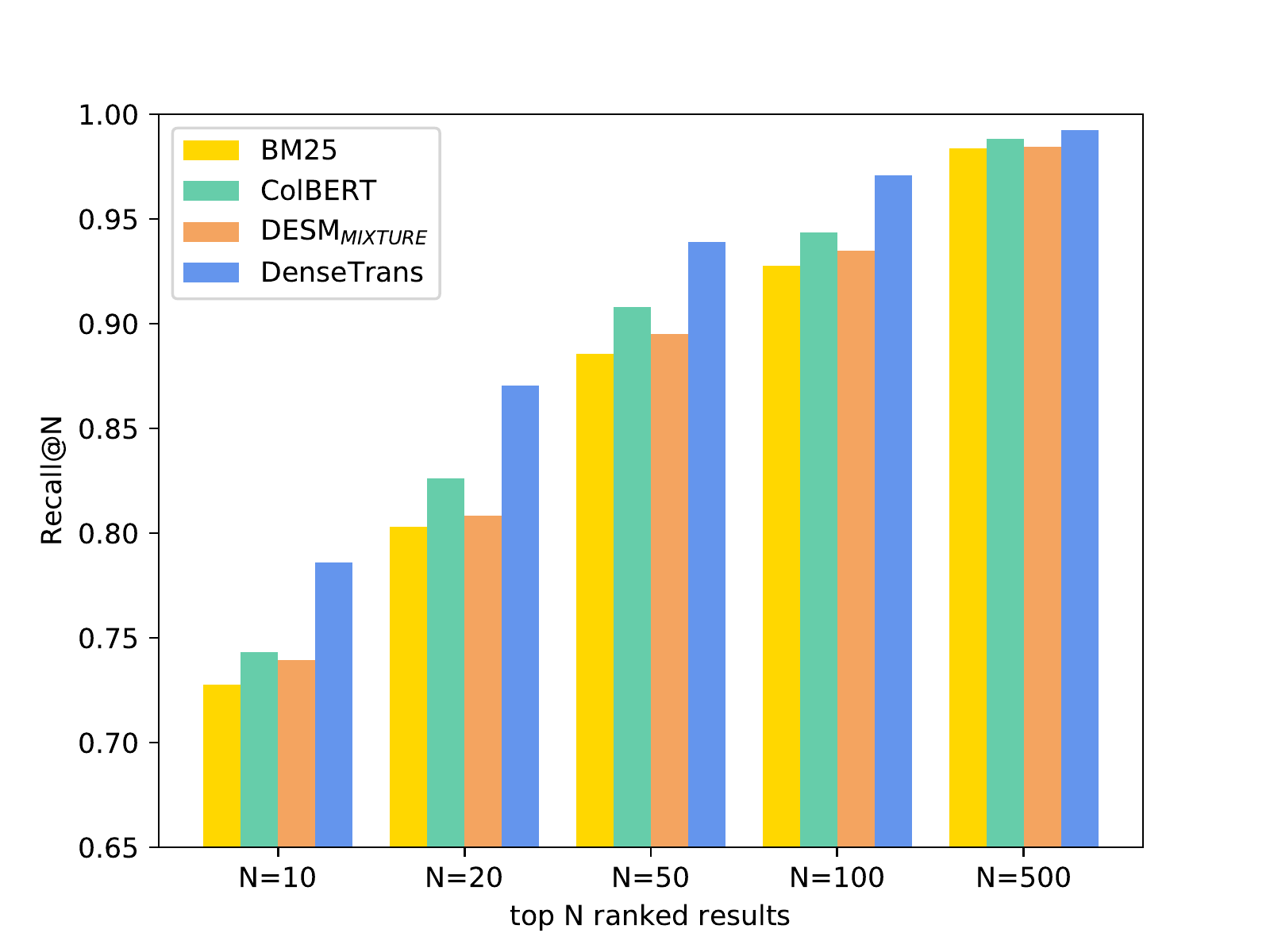}
\caption{The results of Recall under different cutoffs of the retrieval results on Quora dataset.}
\label{fig:topN}
\end{figure}

\subsubsection{Impact of Different Cutoffs of Ranked List}
\label{Different Threshold}
Since the first-stage ranker is to recall a number of potentially relevant texts for latter re-ranking stages, the number of returned texts could vary in different retrieval scenarios. Thus, we analyze the performance of each model for different cutoffs of top ranked questions. Specifically, we compare the DenseTrans model with three strong baselines, including BM25, ColBERT, and DESM$_\mathit{MIXTURE}$. 
The results on Quora dataset are shown in Figure \ref{fig:topN}. 
It can be observed that the DenseTrans model consistently outperforms other methods on all cutoff values. Moreover, it is interesting to see that the DenseTrans model leads to a larger margin compared with other models when retrieving a small number of questions, e.g., the improvement of DenseTrans over BM25 in terms of Recall@10 is $5.8\%$. This is very useful for retrieval tasks as we often focus on a small number of top ranked results, especially on devices with limited resources for re-ranking.

\subsection{Case Study}
To facilitate a better understanding to our proposed model, we perform a case study on Quora dataset. Specifically, we present an example of ranked results of BM25, DenseTrans and DenseTrans$^{-Dense}$ in Table \ref{tab:case_study2}. Due to page limitations, we only show the top-5 ranked results. Here, the DenseTrans$^{-Dense}$ is constructed by removing all the dense connections from the DenseTrans model. 

As is shown in the Table \ref{tab:case_study2}, the input question is ``Who will win the U.S.A presidential elections of 2016?'', which has several keywords, such as ``the U.S.A.'', ``2016'', and ``presidential elections''.
The ground truth question is ``Who will win this US presidential elections 2016?'', which includes the three key elements.
It is interesting to see that BM25 and DenseTrans capture the detailed low-level features as they recall the ground truth question in the top-5 ranked results.
However, the DenseTrans$^{-Dense}$ fails to retrieve the ground truth question in top results by losing the low-level detailed features in the output representations. What's worse, it recalls "Tamil Nadu elections" rather than "the U.S.A presidential elections" in the 5th ranked result. This demonstrates that a highly abstract representation could generalize to the semantic related questions, but could also introduce noise. 
By equipping the DenseTrans$^{-Dense}$ model with dense connections, the DenseTrans model is able to capture both the detailed low-level features as well as the abstract high-level features. Overall, these results provide a vivid explanation that the dense connectivity is indeed helpful to strengthen the discriminative power by involving the detailed low-level features in the learned representations.

\begin{table}[!t] \small
\centering
\caption{An example of top-5 ranked results from BM25 (upper), DenseTrans (middle) and DenseTrans$^{-Dense}$ (bottom) models on Quora dataset. The ground-truth question in retrieval results is marked with red color.}
\label{tab:case_study2}
\begin{tabular}{c|l}
\hline
question & Who will win the U.S.A presidential elections of 2016? \\
ground & Who will win this US presidential elections 2016?  \\
\hline
\#1 & {\color{red} Who will win this US presidential elections 2016?} \\
\#2 & Who will win the 2016 presidential elections?\\
\#3 & Who will win the presidential election of 2016?\\
\#4 & Who will win American Election in 2016?\\
\#5 & Who will win the US elections 2016?\\
\hline
\#1 & Who will win the 2016 presidential elections?\\
\#2 & Who will win the US elections 2016?\\
\#3 & {\color{red} Who will win this US presidential elections 2016?} \\
\#4 & Who will win in America presidential elections in 2016?\\
\#5 & Who will win the 2016 U.S. presidential election and why?\\
\hline
\#1 & Who do you think will be the next US president?\\
\#2 & Who is the next US president in your think?\\
\#3 & Who do you think is going to win the presidential elections... \\ 
\#4 & Who will be the next President of America and Why?\\
\#5 & Who will win Tamil Nadu elections 2016?\\
\hline
\end{tabular}
\end{table}

\section{Conclusions and Future Work}
In this paper, we propose to address the vocabulary mismatch problem for the first-stage retrieval in question retrieval task.
To satisfy the semantic requirement and the discriminative requirement in building the first-stage ranker, we propose a novel DenseTrans model. The DenseTrans model learns standalone semantic representations for question pairs with a stack of several Transformer layers. Moreover, we introduce the dense connectivity between the Transformer layers to strengthen the discriminative power during semantic representations abstracting.
Experimental results on Quora and WikiAnswers datasets show that the DenseTrans model outperforms strong term-based, embedding-based and hybrid methods on all the evaluation metrics. 

For future work, we would like to further investigate the extent of the requirements on semantics than on discrimination. Also, we are interested in applying DenseTrans to other challenging retrieval tasks, such as ad-hoc retrieval and answer retrieval.

\begin{acks}
This work was funded by Beijing Academy of Artificial Intelligence (BAAI) under Grants No. BAAI2019ZD0306, the National Natural Science Foundation of China (NSFC) under Grants No. 62006218, 61902381, 61773362, and 61872338, the Youth Innovation Promotion Association CAS under Grants No. 20144310, 2016102, and 2021100, the Lenovo-CAS Joint Lab Youth Scientist Project, and the Foundation and Frontier Research Key Program of Chongqing Science and Technology Commission (No. cstc2017jcyjBX0059).
\end{acks}

\normalem
\bibliographystyle{ACM-Reference-Format}
\bibliography{paper}

\end{document}